\documentclass[prd,aps,preprintnumbers,nofootinbib,twocolumn]{revtex4}
\usepackage{epsfig}
\usepackage{amsmath}
\usepackage{hyperref}

\begin{document}
	
	\preprint{Cavendish-HEP-24/01}
	\preprint{IFJPAN-IV-2024-13}
	\preprint{P3H-24-087}
	\preprint{TTK-24-48}
	
	\renewcommand{\thefigure}{\arabic{figure}}
	
	\title{Open $B$-hadron production at hadron colliders in QCD at next-to-next-to-leading-order and next-to-next-to-leading-logarithmic accuracy}
	
	\author{Micha\l{}  Czakon}
	\affiliation{{\small Institut f\"ur Theoretische Teilchenphysik und Kosmologie, RWTH Aachen University, D-52056 Aachen, Germany}}
	\author{Terry Generet}
	\author{Alexander Mitov}
	\affiliation{{\small Cavendish Laboratory, University of Cambridge, Cambridge CB3 0HE, UK}}
	\author{Rene Poncelet}
	\affiliation{{\small The Henryk Niewodnicza\'nski Institute of Nuclear Physics, ul. Radzikowskiego 152, 31-342 Krakow, Poland}}
	
	\date{\today}
	
	\begin{abstract}
		We report on a calculation of open heavy-flavor production at hadron colliders which extends to next-to-next-to-leading order (NNLO) accuracy the classic NLO-accurate formalism developed almost 30 years ago under the acronym FONLL. The approach retains the exact heavy-flavor mass dependence at low transverse momentum, $p_T$, and resums collinear logarithms through next-to-next-to-leading log (NNLL) at high $p_T$. Provided are predictions for $B$-hadrons as well as $B$-decay products like $J/\Psi$ and muons.  The main features of the NNLO+NNLL results are reduced scale dependence and moderate NNLO correction, consistent with perturbative convergence in a wide range of kinematic scales from few GeV up to asymptotically large values of $p_T$. The new calculation significantly improves the agreement with data for $B$-hadrons and muons. We uncover an intriguing discrepancy in $J/\Psi$ final states which may point to a lower value of the $B\to J/\Psi$ decay rate.
	\end{abstract}
	\maketitle

	\section{Introduction}

	The inclusive production of identified heavy-flavored hadrons at hadron colliders is a process of special interest. First of all, this process has one of the highest production rates. Furthermore, due to the mass $m$ of the underlying heavy quark being sufficiently larger than $\Lambda_{\rm QCD}$, the process can reliably be described with perturbative methods. The precise theoretical description is essential in this case because it influences all measurements involving heavy hadron decays which, in turn, are a window to understanding Standard Model (SM) physics and many searches for new physics. Extensive heavy-flavor production measurements exist, starting almost four decades ago at the CERN $Sp\bar pS$ collider and continuing through the Tevatron and LHC programs. The measurements span an impressive kinematic range and various hadronic states. 
	
	The modern foundation for the theoretical description of the process goes back to the classic work of Mele and Nason \cite{Mele:1990cw} who introduced the so-called perturbative fragmentation function (PFF) formalism. Combining this formalism with the then-newly-developed next-to-leading order (NLO) QCD calculation \cite{Nason:1989zy,Beenakker:1990maa} of differential heavy-flavor production, and the massless NLO coefficient functions \cite{Aversa:1988vb}, the so-called FONLL formalism emerged \cite{Cacciari:1998it}. It is the most widely used description of fragmentation processes with heavy flavors at hadron colliders, see for example refs.~\cite{Cacciari:2002pa,Cacciari:2003uh,Cacciari:2005rk,Cacciari:2012ny,Cacciari:2015fta}. Alternative approaches have also been developed and are actively explored \cite{Aivazis:1993pi,Collins:1998rz,Kniehl:2004fy,Kniehl:2005mk,Guzzi:2011ew,Kusina:2013slm,Helenius:2018uul}.
	
	Heavy-flavor production at low transverse momenta, $p_T\sim m$, is reliably described in fixed order perturbation theory where the heavy flavor is neither included in the running of the strong coupling constant $\alpha_S$ nor in the evolution of the parton distribution functions (PDFs). In the opposite kinematic regime, $p_T\gg m$, the heavy flavor is essentially massless and is included in $\alpha_S$ running and PDF evolution. At such high scales fixed order perturbation theory in terms of $\alpha_S$ breaks down due to the appearance of large collinear logs $\log(p_T/m)$ to all orders in the coupling constant. The FONLL approach provides a unified description of open heavy-flavor production for scales as low as the mass of the heavy flavor, $p_T\sim m$, and as high as the measurements at colliders allow, i.e.~hundreds of GeV or even more, where $p_T\gg m$. 

	\begin{figure}[t]
		\includegraphics[width=0.40\textwidth]{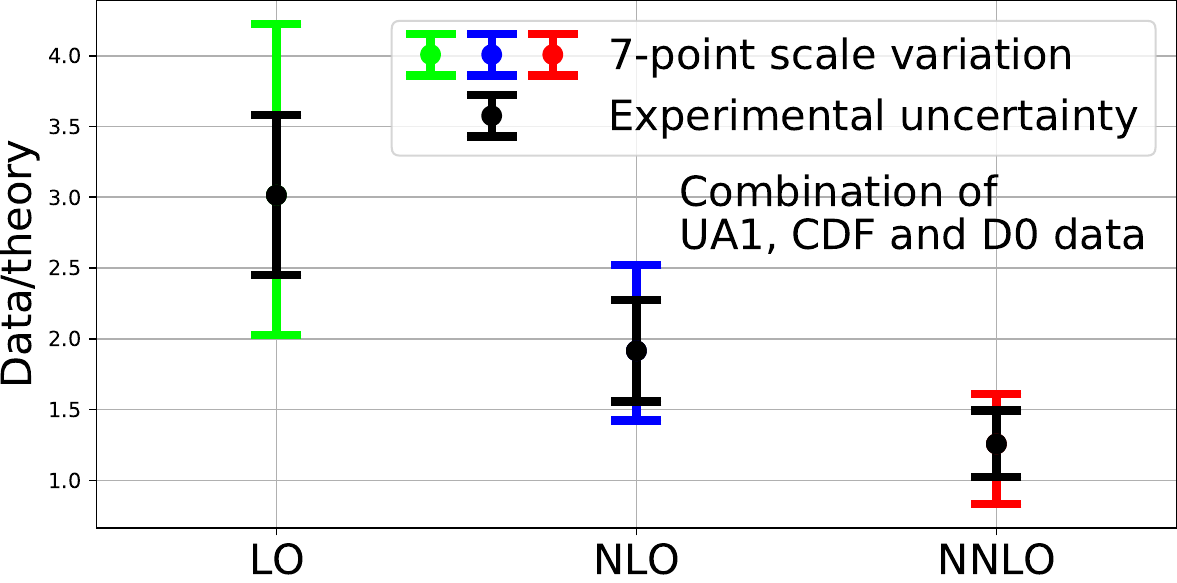}
		\caption{Fiducial $\bar b b$ cross section in LO, NLO and NNLO QCD. The data uncertainty is based on a naive combination of ten separate measurements \cite{UA1:1985bnq,UA1:1990vvp,UA1:1993jok,CDF:1992sue,CDF:1993fax,CDF:1993dog,CDF:1994zxd,CDF:1996kan,D0:1999hay,CDF:2007pxr}.}
		\label{fig:Tevatron}
	\end{figure}

	Despite the major improvement achieved by the FONLL formalism, its predictions still suffer from large theoretical uncertainty from missing higher-order terms. This is especially the case at low $p_T$ since $\alpha_S(p_T\sim m)$ is large and the perturbative expansion converges slowly. For this reason it has been suspected for a very long time that the inclusion of higher order corrections beyond NLO is needed. Another reason driving the need for NNLO QCD corrections is that past comparisons between data, especially from the Tevatron, and theory predictions showed very large discrepancy. Indeed, early theory/data comparisons showed differences as large as a factor of 3, while the inclusion of the full FONLL formalism reduced this discrepancy by about a half. This can be seen in fig.~\ref{fig:Tevatron} where, in anticipation of our main results, one can see that higher order corrections indeed significantly modify the theory predictions, and the inclusion of NNLO corrections finally leads to a theory/data agreement.
	
	Calculations of open bottom production at NNLO have already appeared in the literature. The first such calculation was performed in ref.~\cite{Catani:2020kkl} where predictions were obtained in fixed-order perturbation theory, without resummation, and at the level of $b$ quarks, i.e.~no hadronization effects were included. Subsequently, ref.~\cite{Mazzitelli:2023znt} provided a first calculation at NNLO matched to a parton shower (PS). This calculation provides NNLO$_{\rm PS}$-accurate predictions for $B$ hadrons with resummation at leading logarithmic accuracy. Our present work differs from these calculations by including both the relevant hadronization effects and the resummation of collinear logs at full NNLL accuracy. As we demonstrate in the following, the effect of the resummation with high-logarithmic accuracy leads to major improvement in the theory predictions and has a very significant impact at large $p_T$.

	\section{Structure of the NNLO+NNLL prediction}

	As already alluded to in the previous section, there are two distinct approaches to calculating open $B$ production: we call one of them {\it massive} and the other one {\it massless}. In each case one calculates the suitably defined partonic cross-section for a respectively massive or massless $b$ quark, which is then modified to account for the $b\to B$ fragmentation. 
	
	The differential cross-section for producing a $B$ hadron in the massive approach is denoted as $\sigma_B(m,p_T,\mu)$, where $m$ is the on-shell mass of the $b$ quark and $\mu$ denotes the renormalization, factorization and fragmentation scales. The partonic cross section for producing an on-shell massive $b$ quark is computed through NNLO in 4-flavor, $n_f=4$, QCD in complete analogy with the calculation of differential top-quark pair production of ref.~\cite{Czakon:2015owf}, where $n_f = 5$. We use the {\tt STRIPPER} framework \cite{Czakon:2010td,Czakon:2011ve} as implemented in refs.~\cite{Czakon:2014oma,Czakon:2019tmo}. The two-loop amplitudes are taken from ref.~\cite{Chen:2017jvi}. All one-loop amplitudes are obtained with the help of the {\tt OpenLoops 2} library \cite{Buccioni:2019sur} while all tree-level amplitudes are taken from the {\tt AVHLIB} library \cite{Bury:2015dla}.
	
	The fragmentation $b\to B$ is modeled with the fragmentation function (FF) $D_{b\to B}(z)$ extracted in ref.~\cite{Czakon:2022pyz}. It is assumed that all $B$-hadron species admit identical FFs, with fragmentation fractions as measured by LEP \cite{ParticleDataGroup:2024cfk}. The partonic fraction $z$ is well defined only in the strict collinear limit where the $b$ quark is massless. In such a case one has $p_B = z p_b$, where $p_{b,B}$ are the corresponding four momenta. When the $b$ quark is massive, $m\neq 0$, we use the following rescaling prescription:
	\begin{eqnarray}
		\vec{p}_B &=& z \vec{p}_b\,,\nonumber\\
		E_B &=& \sqrt{\vec{p}_B^2+m_B^2}\,,\nonumber
	\end{eqnarray}
	i.e.~three-momenta are rescaled uniformly while energies are fixed by requiring final state particles to be on-shell.
	
	Subsequent decays of the $B$ hadron are treated as follows. For massive-$b$ fragmentation, the decay of the $B$-hadron is treated as an isotropic decay in the $B$-hadron rest frame. Distributions of $B$ decaying to the final state of interest are obtained from {\tt EvtGen} \cite{Lange:2001uf}, and are then boosted to the lab frame. In contrast, in the massless-$b$ calculation, following ref.~\cite{Czakon:2022pyz}, $B$ decays are included directly in the fragmentation step as an additional convolution on top of the $b\to B$ FF.
	
	The differential cross-section for producing a $B$ hadron in the massless approach is denoted by $\sigma_B(0,p_T,\mu)$. We stress that the name massless is a bit of a misnomer since $\sigma_B(0,p_T,\mu)$ correctly accounts for the leading power in $m$ in the limit $m\to 0$, including both mass-independent and $\log^n(m)$ terms. The name {\it massless} originates in the fact that it is based on massless calculations: the differential massless-$b$ partonic cross section is computed in $n_f=5$ massless QCD, and final state collinear singularities are subtracted numerically in the $\overline{\rm MS}$ scheme as explained in ref.~\cite{Czakon:2021ohs}. The required two-loop four-parton massless amplitudes are taken from ref.~\cite{Broggio:2014hoa}, also computed in refs.~\cite{Bern:2002tk,Bern:2003ck,Glover:2003cm,Glover:2004si,DeFreitas:2004kmi}, while one loop ones are again taken from the {\tt OpenLoops 2} library.
	
	The massive and massless calculations have different ranges of applicability: $\sigma_B(m,p_T,\mu)$ describes $B$ production at low and intermediate $p_T$, while $\sigma_B(0,p_T,\mu)$ is applicable for intermediate and large $p_T$. Ideally, one would like to have a unified prediction, denoted as $\bar\sigma_B(p_T,\mu)$,  which can describe the complete kinematic range. This problem was solved within the so-called FONLL framework. In this work we follow its logic and combine these two calculations as follows:
	\begin{eqnarray}
		\label{eq:best}
		\bar\sigma_B(p_T,\mu) &=& \sigma_B(m,p_T,\mu) \\
		&& + G(p_T)\bigg(\sigma_B(0,p_T,\mu) - \sigma_B(0,p_T,\mu)\Big\vert_{\rm FO}\bigg)\,,\nonumber
	\end{eqnarray}
	where
	\begin{equation}
		G(p_T) = \frac{p_T^2}{p_T^2+c^2m^2}\,,
		\label{eq:G}
	\end{equation}
	and $c=5$ for $B$-hadron final states, $c = 3$ for $J/\Psi$ final states and $c = 0$ (i.e.~no suppression) if the observed final state is a muon. The $p_T$ in the above equation is the $p_T$ of the corresponding final state particle. 
	
	The constant $c$ is modeled after ref.~\cite{Cacciari:1998it}, where the value $c=5$ was chosen for final-state $b$-quarks. Since the difference between the $b$-quark and $B$-hadron momentum is typically small, $c=5$ can also be used for $B$-hadrons. However, $B$-hadron decay products require smaller values for $c$ due to their significantly reduced momentum relative to the fragmenting $b$-quark. We find that the values given above are approximately equivalent to using $c=5$ for $b$-quarks.

	\begin{figure}[t!]
		\includegraphics[width=0.45\textwidth]{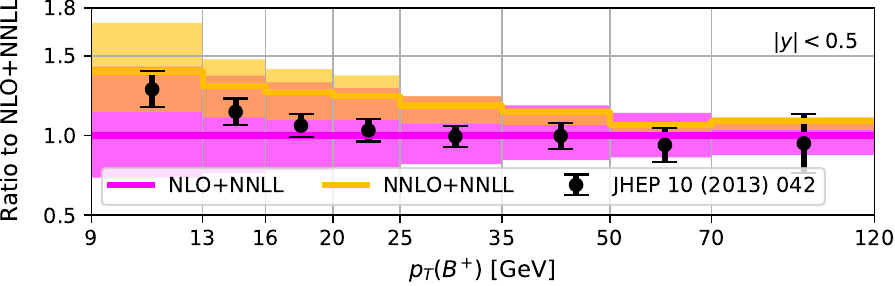}
		\caption{$B^+$ $p_T$ spectrum (see supplementary materials \cite{SM} for more rapidity slices). Shown is our best prediction at NNLO+NNLL (orange) versus the current state of the art NLO+NNLL (purple) versus LHC data. Shown are the ratios w.r.t.~the central prediction at NLO+NNLL.}
		\label{fig:B-LHC-2}
	\end{figure}

	The massless calculation $\sigma_B(0,p_T,\mu)$ is not reliable in the low $p_T$ region. The reason for this is that $\sigma_B(0,p_T,\mu)$ diverges at low $p_T$ and in the region $p_T\sim m$ this divergence is not well compensated for by its fixed order expansion $\sigma_B(0,p_T,\mu)\big\vert_{\rm FO}$. The massless calculation can therefore have a significant effect on the prediction at low $p_T$ where one expects the massive calculation to be reliable. 
	
	This unwanted behavior of the massless calculation $\sigma_B(0,p_T,\mu)$ is suppressed by the function $G(p_T)$ and by two additional modifications on $\sigma_B(0,p_T,\mu)$. 
	First, following ref.~\cite{Cacciari:1998it}, we remap the $p_T$ of the final state particle using $p_T\to \sqrt{p_T^2-m^2}$, vetoing events with $p_T<m$ before any rescaling. In effect, this modification renders the massless calculation finite for $p_T\to 0$ and ensures that the massless calculation does not contribute below the physical partonic center of mass energy of $2m$. 
	
	Second, in the massless calculation we have excluded fixed order terms of order higher than the fixed order accuracy of the calculation. For example at NNLO, the product of the NNLO partonic cross-section is multiplied only by the LO perturbative FF but not by its higher order terms at NLO or NNLO \cite{Melnikov:2004bm,Mitov:2004du}.
	
	Soft-gluon resummation of the heavy quark FF is performed as described in ref.~\cite{Czakon:2022pyz}. DGLAP evolution is performed using the APFEL library \cite{Bertone:2013vaa}. Soft-gluon resummation and DGLAP evolution are always performed at full NNLL.
	
	In both massive and massless calculations we use
	\begin{equation}
		\mu^2 = m^2 + d^2p_T^2\,,
		\label{eq:scale}
	\end{equation}
	with $d=1$ for $b$-quark or $B$-hadron final states, $d = 1.5$ for $J/\Psi$ final states and $d = 2$ for muons. The constant factor $d$ is introduced in order to approximate, on average, the $p_T$ of a fragmenting $b$ quark from the $p_T$ of the final state particle. The value of $p_T$ used in the function $G(p_T)$ in eq.~(\ref{eq:G}), and in the scale $\mu$ in eq.~(\ref{eq:scale}) -- when that scale is used in the massless calculation and its truncation, is the one obtained after the remapping described above. Scale variation around the central scale in eq.~(\ref{eq:scale}) is performed using the 15 point approach. In all cases, the strong coupling constant $\alpha_S$ is taken at NNLL and $m = 4.92$ GeV, both being consistent with the parton distribution set {\tt NNPDF3.1} \cite{NNPDF:2017mvq} we use. In contrast to the FFs -- which are evaluated at different orders in different parts of the calculation as explained above -- we use NNLO parton distribution functions throughout the calculation.

	\begin{figure}[t!]
		\includegraphics[width=0.45\textwidth]{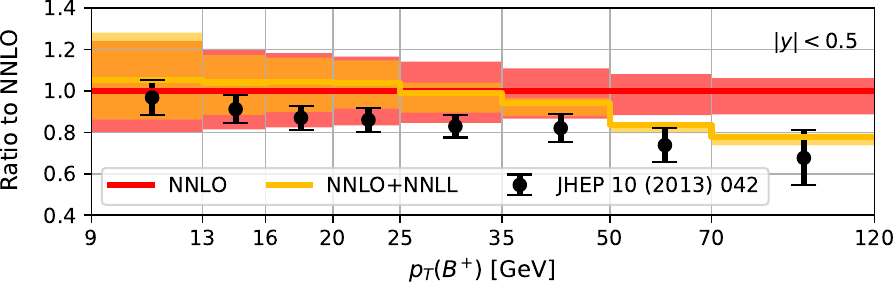}
		\caption{As in fig.~\ref{fig:B-LHC-2} but with NNLO (red) instead of NLO+NNLL.}
		\label{fig:B-LHC}
	\end{figure}

	\section{LHC Phenomenology}
	
	The results derived in this paper allow us to address a number of outstanding questions, in particular the level of agreement between SM predictions and hadron collider measurements of $B$ hadron final states, and the behavior of theoretical predictions at higher perturbative orders across vast kinematics ranges. In order to make our analysis as conclusive as possible we have placed special emphasis on numerical optimization. As a result, all presented predictions are of very high quality, i.e.~the so called Monte Carlo integration error has been reduced to the point that in most bins it is significantly smaller than the dominant scale uncertainty.
	
	While in the following we only present results for relatively central kinematics, we have checked that the pattern of perturbative corrections is almost identical for the forward kinematics relevant for the LHCb experiment.
	
	We first turn our attention to $B^+$-meson production at the LHC. In fig.~\ref{fig:B-LHC-2} we compare our best prediction, eq.~(\ref{eq:best}), computed with NNLO+NNLL accuracy, with the current state of the art at NLO+NNLL (which we take as a proxy for FONLL). Both calculations are compared to 7 TeV ATLAS data \cite{ATLAS:2013cia}. 
	
	At the lowest $p_T$ available, the NNLO+NNLL scale variation is only slightly reduced relative to the NLO+NNLL one. This is not unexpected given the slower convergence of perturbation theory at low scales. 
	The inclusion of the NNLO+NNLL correction nevertheless plays an important role in this kinematic range since it generates a significant shift in the central prediction which brings it closer to data. 

	\begin{figure}[t]
		\includegraphics[width=0.45\textwidth]{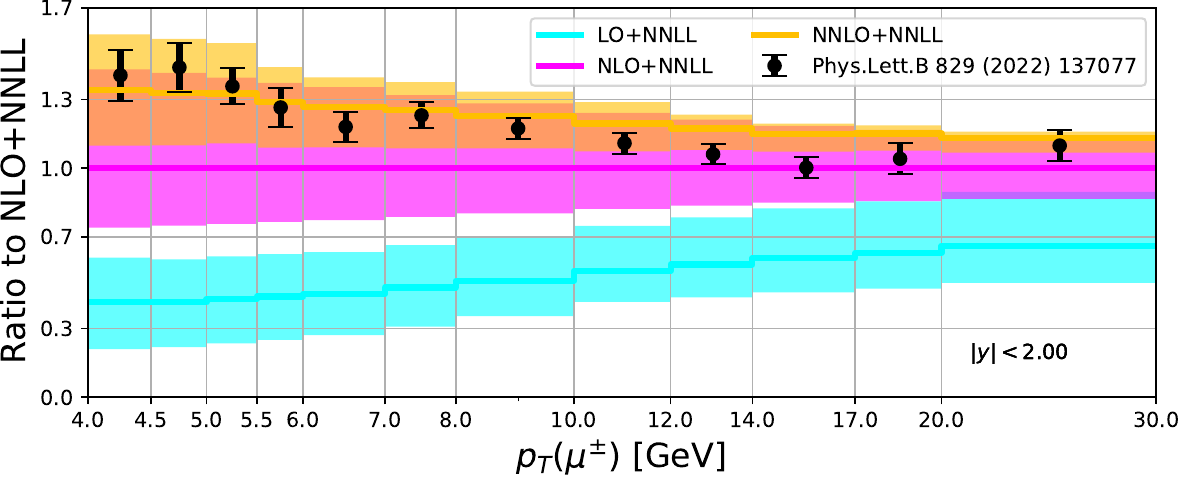}
		\caption{Muon $p_T$ spectrum at LO+NNLL (light blue), NLO+NNLL (purple) and NNLO+NNLL (orange) compared with LHC data.}
		\label{fig:mu}
	\end{figure}

	The uncertainty of the NNLO+NNLL prediction quickly decreases at higher $p_T$. At the highest available $p_T$ it is about a factor of three smaller than the NLO+NNLL one. Such a strong decrease of scale variation indicates that the NNLO+NNLL theory prediction at high $p_T$ is very precise. Indeed, the theory uncertainty in that range is significantly smaller than that of the available data. The combination of normalization and shape changes in the NNLO+NNLL prediction relative to the NLO+NNLL one brings our best prediction into good agreement also with high $p_T$ data. This is true for all rapidity slices. 
	
	Finally, as can be seen in fig.~\ref{fig:B-LHC-2}, the ratio NNLO+NNLL to NLO+NNLL is relatively small in the sense that it is always within the NLO+NNLL scale variation band. This indicates that the theory prediction for this observable at this order of perturbation theory stabilizes and reliably estimates missing yet-higher order corrections.

	\begin{figure}[t!]
		\includegraphics[width=0.45\textwidth]{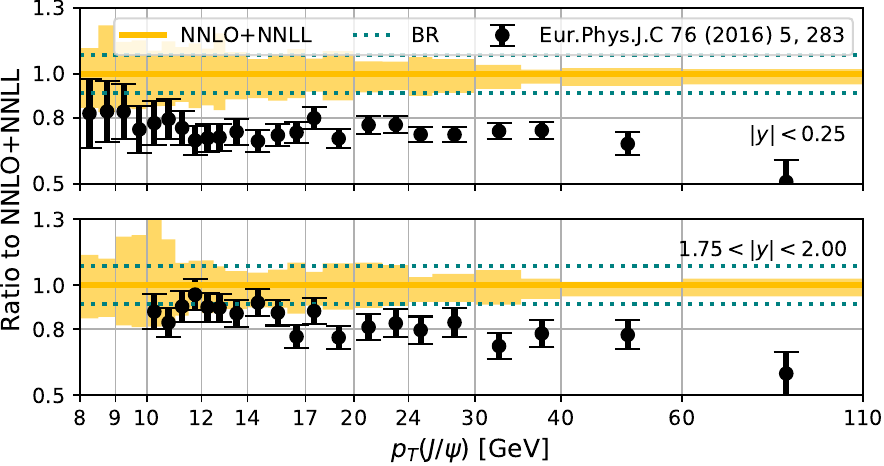}
		\caption{$J/\Psi$ $p_T$ spectrum at NNLO+NNLL (orange) compared with LHC data for two rapidity slices (see supplementary materials \cite{SM} for the other slices). The dotted lines (green) show the uncertainty from $BR(B\to J/\Psi)$.}
		\label{fig:J/Psi}
	\end{figure}

	Another important comparison is shown in fig.~\ref{fig:B-LHC} where the NNLO massive fixed-order prediction is compared with the NNLO+NNLL one. Both predictions are compared to the data already shown in fig.~\ref{fig:B-LHC-2}. As expected, at low $p_T$ the two agree very well (which, in a way, implies that the matching eq.~(\ref{eq:best}) of massive and massless calculations works well). As the $B$-hadron $p_T$ increases, however, the resummed calculation's scale variation decreases quickly and the resummed prediction becomes significantly more precise than the fixed order one. This behavior is again expected in principle, however, it is impressive to see that the improvement at the highest available $p_T$ is as large as a factor of three.
	
	Even more interestingly, for $p_T \gtrsim 50$ GeV or so, the two predictions do not overlap any more and their difference becomes more and more pronounced as $p_T$ increases. The difference between the two predictions is significant relative to the size of the two uncertainty bands. Fig.~\ref{fig:B-LHC}, therefore, offers an unambiguous and affirmative answer to the often asked question if high-$p_T$ resummation effects are important in this class of observables. We would like to stress that the confidence with which one can draw this conclusion is underpinned by the much reduced scale uncertainty at NNLO, and especially at NNLO+NNLL, and that with NLO accuracy one would not have been able to draw such a conclusion. 
	
	We next turn our attention to muon final states originating in semileptonic $B$ decays. In fig.~\ref{fig:mu} we show the muon $p_T$ distribution and compare it to recent 5.02 TeV ATLAS data \cite{ATLAS:2021xtw}. We show LO, NLO and NNLO predictions all supplemented with NNLL resummation. The pattern of higher order corrections is broadly similar to the case of $B^+$-mesons and shows large NLO/LO correction and a relatively moderate NNLO/NLO one. Notably, the NNLO+NNLL uncertainty band almost completely overlaps with the NLO+NNLL one which strongly indicates that at this order of perturbation theory the theory prediction for this observable becomes reliable. Another notable feature is the size of the scale uncertainty. The inclusion of the NNLO correction reduces the scale variation at the lowest available $p_T$ almost in half, while at high $p_T$, similarly to the case of $B$ hadrons, the scale uncertainty is reduced by over a factor of three. 
	
	The NNLO correction affects both the normalization and shape of the muon $p_T$ spectrum, and brings it very close to data in the full $p_T$ range. While the NLO+NNLL prediction is also consistent with data within its much larger uncertainty band, it is clear that the shapes of the NLO+NNLL prediction and data are not perfectly aligned. The NNLO+NNLL prediction agrees quite well with the overall shape of the measured $p_T$ spectrum despite its much reduced scale uncertainty band. Clearly data is much more precise than theory for low values of the muon $p_T$, however this changes at higher $p_T$ where the data uncertainty becomes the dominant one. 
	
	Finally, in fig.~\ref{fig:J/Psi} we show the $p_T$ distribution of $J/\Psi$ final states originating in $B$ decays. Several rapidity slices are considered. The overall pattern of higher order corrections, not shown here, is similar to the ones observed for $B$ final states, including the pronounced difference at high $p_T$ between the fixed order massive prediction and the resummed one, as well as the significant reduction of scale uncertainty at high $p_T$ due to resummation. The comparison with 8 TeV ATLAS data \cite{ATLAS:2015zdw}, however, is at variance with the overall excellent agreement observed for $B$ and $\mu$ final states. As can be seen in fig.~\ref{fig:J/Psi}, the theory/data agreement is not satisfactory. Although not shown here, we have checked that the theory/data comparison for $\Psi(2S)$ final states shows similar level and pattern of disagreement. 
	
	The much reduced theory uncertainty at NNLO+NNLL, which is comparable to the one of the data for all datapoints, allows us to draw quantitative conclusions about the origin of this clear discrepancy.
	
	The overall excellent theory/data agreement for $B^+$ and $\mu$ final states makes it unlikely that this discrepancy can be attributed to the underlying theoretical description -- including our general treatment of $B$ decays. In our view, the most likely origin of this theory/data incompatibility is the $B\to J/\Psi$ branching ratio (BR).
	
	The BR used in fig.~\ref{fig:J/Psi} is taken from the Particle Data Group (PDG) \cite{ParticleDataGroup:2024cfk}, $BR(B\to J/\Psi) = (1.16\pm 0.10)\cdot10^{-2}$, which corresponds to an admixture of $B$ hadrons. Somewhat surprisingly, this BR is known with a relatively low accuracy of about 9\%. This BR uncertainty is shown in fig.~\ref{fig:J/Psi} with the two green dotted lines. In our current context, such an uncertainty is significant and, in fact, is the dominant one for most datapoints, surpassing both the experimental and theory scale uncertainties. While a detailed analysis of this theory/data discrepancy is beyond the scope of this publication, we note that, should the $BR(B\to J/\Psi)$ be lower than the PDG value -- perhaps by about 25\% or so -- theory and data would become in agreement for almost all $p_T$ and rapidity values.

	\section{Summary and Outlook}
	
	We extend the FONLL formalism for open heavy-flavor production at hadron colliders to NNLO+NNLL accuracy. Our result exhibits a much reduced scale uncertainty, especially at moderate and large $p_T$. Comparisons with LHC data show excellent agreement for $B$-hadrons and muons from $B$ decays. A notable deviation for $J/\Psi$ final states is observed, which may be due to the value of $BR(B\to J/\Psi)$. Our work offers the possibility for precisely constraining various parameters related to $B$ production and decay and may impact important ongoing LHC $B$-physics measurements.

	\begin{acknowledgments}
		We would like to thank Jonas Lindert for providing certain one-loop amplitudes. R.P.~wishes to thank the CERN Theoretical Physics Department and the Aspen Center for Physics, which is supported by National Science Foundation grant PHY-2210452 and by a grant from the Simons Foundation (1161654, Troyer), for its hospitality while part of this work was carried out. The work of M.C.~was supported by the Deutsche Forschungsgemeinschaft (DFG) under grant 396021762 - TRR 257: Particle Physics Phenomenology after the Higgs Discovery. T.G.~and A.M.~have been supported by STFC consolidated HEP theory grants ST/T000694/1 and ST/X000664/1. This work was performed using the Cambridge Service for Data Driven Discovery (CSD3), part of which is operated by the University of Cambridge Research Computing on behalf of the STFC DiRAC HPC Facility (www.dirac.ac.uk). The DiRAC component of CSD3 was supported by STFC grants ST/P002307/1, ST/R002452/1 and ST/R00689X/1. Simulations were also performed with computing resources granted by RWTH Aachen University under project p0020025.
	\end{acknowledgments}


\begin{thebibliography}{99}
\bibitem{SM}
		See the ancillary files for versions of figures 2, 3 and 5 that show all rapidity slices.
\end{thebibliography}
\end{document}